\documentclass[useAMS,usenatbib,fleqn]{mn2e}
\usepackage{amsmath,amssymb,amsfonts,latexsym}
\usepackage{longtable}
\usepackage{fixltx2e}
\usepackage{lscape}
\usepackage{times}
\usepackage[pdftex]{graphicx}
\def\apj{{ApJ}}

\def\apjl{{ApJL}}

\def\aap{{A\&A}}

\def\mnras{{MNRAS}}
\def\aaps{{Astron. Astrophys. Suppl.}}

\def\nat{{Nature}}

\def\prd{{Physical Review D}}
\def\pre{{Physical Review E}}
\def\apss{{Ap\&SS}}
\def\04a{{2004 a}}
\def\04b{{2004 b}}
\title{Radio pulsar activity and the crustal Hall drift}

\author[U.~Geppert \& J.~Gil \& G.~Melikidze]{U.~Geppert$^{1,2}$\thanks{E-mail:ulrich.geppert@dlr.de}, J.~Gil$^{1}$, G.~Melikidze$^{1,3}$\\
$^{1}$ Kepler Institute of Astronomy, University of Zielona G\'{o}ra, Lubuska
    2, 65-265, Zielona G\'{o}ra, Poland \\
$^2$ German Aerospace Center, Institute for Space Systems, , Robert-Hooke-Str. 7, 28359 Bremen, Germany\\
$^3$ Abastumani Astrophysical Observatory, Ilia State University, 3-5 Cholokashvili str., Tbilisi, 0160, Georgia\\
}

\begin{document}

\date{Accepted . Received ; in original form }

\pagerange{\pageref{firstpage}--\pageref{lastpage}} \pubyear{2013}

\maketitle

\label{firstpage}
\begin{abstract}
Models of pulsar radio emission that are based on an inner accelerating region require the existence of very strong and
small scale surface magnetic field structures at or near the canonical polar cap. The aim of this paper is to identify
a mechanism that creates such field structures and maintains them over a pulsar's lifetime. The likely physical process
that can create the required 'magnetic spots' is the Hall drift occurring in the crust of a neutron star. It is
demonstrated, that the Hall drift can produce small scale strong surface magnetic field anomalies (spots) on timescales
of $10^4$ years by means of non-linear interaction between poloidal and toroidal components of the subsurface magnetic
field. These anomalies are characterized by strengths of about $10^{14}$ G and curvature radii of field lines of about
$10^6$ cm, both of which are fundamental for generation of observable radio emission.
\end{abstract}


\section{Introduction}
About 45 years after the discovery of the first radio pulsars the physics of the generation of the radio emission is
still a subject of intense scientific debates. There is a general agreement that the strong potential drop above the
polar cap surface must exist to accelerate the charged particles to ultra-relativistic energies. Two pioneering papers
by \citet{RS75} and by \citet{AS79} started an intensive research in two different classes of models: non-stationary
(sparking gap) and stationary (free flow), respectively. Observational and theoretical evidences are searched for and
studied to support or discard one or the other model. The phenomenon of drifting sub-pulses and the combined
observations of pulsars in radio and X-rays \citep{GMZ07a,GMZ07b,GHMGZM08} support strongly the idea that the Partially
Screened Gap  \citep[PSG hereafter,][]{GMG03} is the case, in which the pure vacuum potential drop is partially
screened by thermal iron ions \citep{CR77}.

The creation of a dense electron-positron plasma is the basis of pulsar radio emission processes. These pairs can be
created by photons originating from inverse Compton scattering and/or from curvature radiation, both interacting with a
very strong and curved magnetic field as compared to the global dipolar field $B_d$ at the surface. Therefore, the
small scale field has to be generated in the crust of the neutron star. Already \citet{RS75} in their pioneering work
on the inner acceleration gap noticed implicitly a necessity of small scale surface magnetic field with the radius of
curvature of about $10^6$ cm, much smaller than $10^8$ cm that is typical for the global dipole field. These authors,
however, did not specify a mechanism that could generate such a field structure at the polar cap. In our paper we
explore whether the Hall drift can play that role.

To build up an accelerating potential drop in the polar gap the cohesive energy of charges at the polar cap surface has
to exceed some critical value. This will prevent particles from escaping the stellar surface. As soon as the electric
field exceeds a certain threshold an avalanche pair creation discharges the ultrahigh gap potential drop. The cohesive
energy $\epsilon_c$  is strongly dependent on the surface magnetic field strength $B_s$, which should be much stronger
than $B_d$. As demonstrated by \citet{ML07}, who performed calculations for helium, calcium and iron atoms at the
surface, for each value of the magnetic field $B_s$ there is a a critical temperature $T_s$ above which the
accelerating gap cannot be formed. This critical temperature and corresponding cohesive energy are represented by the
red line in Fig. 1 connecting the data points taken from the paper of \citet{ML07}. No gap can exist above this line
and pure vacuum gap is possible below it. The PSG with some amount of thermally released iron ions can exist in a
narrow region along the red line (see section 2 for more details).

X-ray observations of radio pulsars \cite[see e.g.][]{ZSP05,KPG06,GHMGZM08,PKWG09} indicate that a polar cap surface is
strongly heated due to the bombardment with back flowing charges accelerated within a polar gap. Given the magnetic
field dependence on the cohesive energy, the actual field strength necessary to form a gap has to be much stronger than
global dipolar field $B_d$. These observations estimate the size of the bombarded area ('hot spot') being  less then
$100$ m, suggesting a locally very small scale field. This strong and small scale field anomaly, the 'magnetic spot',
is obviously located within or close to the conventional polar cap area introduced by \citet{GJ69}. The actual polar
cap ('hot spot') is the locus of those anomalous magnetic field lines that join with the open dipolar field lines at
some higher altitudes \citep{GMM02}. The surface area of this actual polar cap is at least one order of magnitude
smaller then the conventional polar cap. The black-body temperature of the 'hot spot' is estimated to exceed one
Million Kelvin, which indicates very strong surface magnetic fields close to or even above $10^{14}$ G (see Fig. 1).

While the PSG model explains quite well the variety of pulsar observations both in radio and X-ray bands, a missing
link is a process that can create and maintain the required strong and small scale surface magnetic field structures
over the lifetime of radio pulsars, i.e. $\sim 10^6 -  10^7$ years. These structures should have high strength $B_s >
5\times 10^{13}$ G to provide high enough cohesive energy to form PSG, and have small radius of curvature
$R_{\text{cur}} \sim 10^6$ cm to cause creation of large enough number of electron-positron pairs to generate the
coherent observable radio emission by means of some kind of plasma instabilities \citep{PM80, AM98, MGP00, GLM04}.

In this paper we will describe in some detail the idea that the Hall drift of the crustal magnetic field could generate
the required surface field structures out of a strong ($\sim 10^{15}$G) toroidal field component, which has been formed
immediately after birth of the neutron star. For simplicity  we assume that it has an axisymmetric geometry. The
currents maintaining this field circulate in deep crustal layers, where the high electric conductivity guarantees
lifetimes well exceeding $10^6$yrs. This mechanism has been proposed recently by \citet{GGMPV12} and will be discussed
here in extenso. The purpose of this paper is not to present a set of models that describe exactly the creation of the
required 'magnetic spots'. On the currently available level of modeling the magneto-thermal evolution (2D computations
only) such description is not yet possible. Rather, we want to introduce a mechanism that is apparently able to
maintain the required magnetic structures over a pulsar lifetime.

In Section 2 basic ideas of the PSG model will be explicated. Observational indications for the size of the magnetic
spot in the polar cap, the corresponding surface field strength there and its surface temperature will be discussed.
Processes and conditions for the creation of electron-positron pairs as well as the estimates for the necessary
magnetic field scales will be described in Section 3. In Section 4. we will discuss possible processes that might
create 'magnetic spots' and argue in favor of the Hall drift. The role of the 'initial' magnetic field configuration
will be examined. The magneto-thermal evolution in the crystallized crust and the obtained results for the creation of
'magnetic spots' are presented in Section 5. These spots are characterized by their surface field strengths and the
curvature radius of their field lines. These characteristic quantities are results of the numerical solution of the
Hall induction equation, assuming state of the art crustal micro-physics, equation of state, and an initial crustal
field configuration that consists of a strong ($\sim 10^{15}$G) toroidal component and a dipolar field  standard for
radio pulsars  ($\sim 10^{12}  -  10^{13}$G). In Section 6 we summarize the results as well as discuss the necessary
extension of the model to non-axial symmetric configurations.

\section{The Partially Screened Gap Model}

The PSG model \citep{GMG03} is an refinement of the vacuum gap model developed by \citet{RS75}. If the
electron-positron pair production really occurs in the inner accelerator potential drop, then one should expect an
intense thermal radiation from the hot polar cap. Indeed, the accelerated positrons will leave the acceleration region,
while the electrons will bombard the polar cap surface, causing heating of the polar cap surface to MK temperatures. At
the same time thermal ejection of iron ions should occur, which will partially screen the polar cap electric field.
This is the essence of the PSG pulsar model. The pure vacuum gap produces too much thermal X-ray emission, in contrast
with observations, so the PSG model is a natural and expected advancement.

In order for the PSG model to function it requires high enough cohesive energy of iron ions to bind them at least
partially at the polar cap surface. We refer to the most recent calculations of cohesive energies by \citet{ML07}
represented by the curved line in Fig.1. As shown by these authors the critical surface temperature $T_i$ , below which
the PSG can form, depends strongly on the actual surface magnetic field $B_s$, that must be close to $10^{14}$ G in all
pulsars for which the surface temperature ($T_s$) of their heated polar cap exceeds $10^6$ K. This is strongly
suggested by X-ray spectral observations of a number of pulsars shown in Fig. 1. Both the critical $T_i$ and the actual
$T_s$ temperatures must be close to each other by means of the thermostatic regulation mechanism. Indeed, if $T_i$ is
slightly higher than $T_s$, then heating by the bombardment becomes more intense, consequently increasing the surface
temperature  thus making heating less intense. This is a classical thermostat mechanism, which will set $T_s \sim T_i$
in PSG along the red line and its nearest vicinity (see discussion in Section 6) presented in Fig. 1.

The parameters of the black body fit to the X-ray spectrum of some pulsars provide very important information for the
understanding of the physical conditions at the neutron star surface. The black body fit allows us to obtain directly
both the 'hot spot' area $A_{\text{BB}}$ and its temperature $T_{\text{BB}}$. In most cases $A_{\text{BB}}$ is much
smaller than the conventional polar cap area $A=6.6\times 10^8 P^{-1}$ cm$^2$, where $P$ is the pulsar period
\citep[see][]{ZSP05, KPG06}. This can be easily explained by assuming that the surface magnetic field of pulsars $B_s$
differs significantly from the surface value of the global dipole $B_d(R)$, where $R$ is the neutron star radius. Then,
one can estimate an actual surface magnetic field by the magnetic flux conservation law as $b = A/A_{\text{BB}} =
B_s/B_d$, where $B_d=2\times 10^{12} (P\dot{P}_{-15})^{0.5}$ G is the dipolar magnetic field at the polar cap surface,
$\dot{P}_{-15} = \dot{P} \times 10^{-15}$ and $\dot{P}$ is the period derivative. Then the value of the actual surface
magnetic field at the hot polar cap can be expressed as

\begin{equation}
B_s=1.3 \times 10^{21} A_{\text{BB}}^{-1} (P\dot{P}_{-15})^{0.5} G, \label{}
\end{equation}

\noindent where $A_{\text{BB}}$ is 'hot spot' area. In most cases $b >> 1$ implying $B_s >> B_d$, while $T_{\text{BB}}
\sim (2 - 4) \times 10^6$ K. This is shown in Fig. 1 , which also confirms observationally the cohesive energy
calculations performed by \citet{ML07}. Indeed, the data points and their errors represent a number of pulsars in which
estimation of the surface area (thus the surface magnetic field) and the surface temperature were possible from X-ray
observations \citep[e.g.][]{S13}. They all lie close to and on the right side of the PSG line in the region where gap
formation is possible. The estimated surface magnetic field $B_s$ exceeds slightly $10^{14}$ G for all data points. The
exemplary values of the dipolar surface magnetic field corresponding to $b = A/A_{\text{BB}} = B_s/B_d$ taken as $100$
are shown on the upper abscissa.

\begin{figure}
\centering
\includegraphics[scale=0.5]{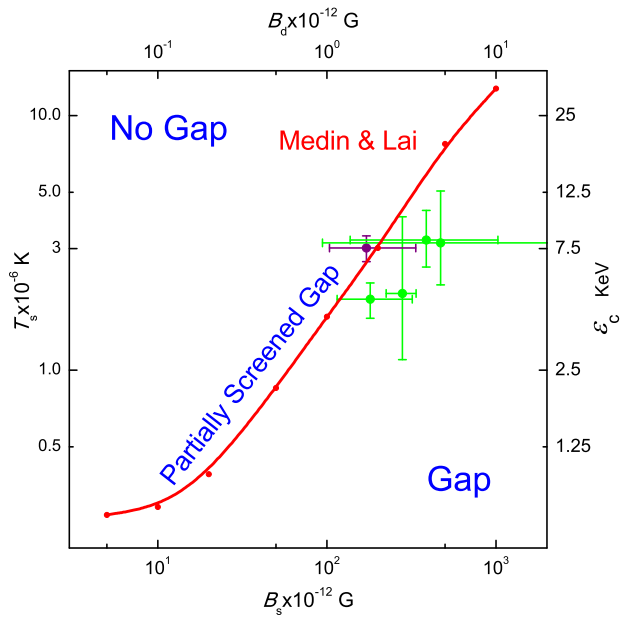}
\caption{Conditions for the creation of a vacuum gap above polar cap surface layers consisting of iron (according to
\citet{ML07}) with given surface temperature $T_s$ and surface magnetic field $B_s$.  For an iron surface temperature
$T_s> 10^6$K, surface field strength of $B_s \gtrsim 10^{14}$G is necessary (for carbon and helium surfaces the
conditions are even more stringent). PSG can be formed in a narrow belt along the red line. Green data points and their
error bars come from various sources \citep[see e.g.][and references therein]{S13}, while the purple dot and
corresponding error bars for PSR B0943+10 are extracted from Table 3 in the newest paper by \citet[][see also
discussion in the text below]{MTET13}.} \label{Q(B,T)}
\end{figure}

For typical radio pulsar parameters with $T_s \gtrsim 2 -  3\times 10^6$K which implies, that a vacuum gap can only be
formed if $B_s$ exceeds $10^{14}$G. Flux conservation arguments  relate the area of the polar cap encircled by the last
open field lines to the black-body area estimated from X-ray observations. \citet{ZSP05} and \citet{KPG06} found $B_s
\approx 4\times 10^{14}$~G, $T_s \approx 3\times 10^6$~K for PSR B1133+16  while its dipolar field derived from $P$ and
$\dot{P}$ is $B_d \approx 4\times 10^{12}$G. Analogously  $B_s \approx 2\times 10^{14}$~G, $T_s \approx 3\times 10^6$~K
for PSR  B0943+10, while for its dipolar component $B_d \approx 6\times 10^{12}$~G at the polar cap. These estimates,
based on combined radio and X-ray observations of isolated neutron stars, indicate the existence of a small scale but
very strong magnetic field $B_s$ that is concentrated in an indeed 'hot spot' within or very close by to the
conventional pulsar polar cap \cite[see Fig. 1 in ][]{SMG10}. Although the photon statistics is rather poor in these
observations, the conclusions about the small sizes and high $T_s$ of these spots are strongly supported by the
theoretical arguments of the balance of cooling and heating by bombardment with ultra-relativistic particles. Moreover,
the recent long time exposure observations of B0943+10 by XMM-Newton \citep{H13} leave no doubt about the existence of
very small hot spots associated with the polar cap of radio pulsars.

Now we can give a good example of PSG model application to a real pulsar. We have just came across the new paper by
\citet{MTET13}, in which the authors reanalyze good quality data taken during simultaneous observations of PSR B0943+10
in radio and X-ray bands \citep{H13}. This old (5 Myr) pulsar with $P=1.1$ s and $\dot P_{-15}=3.5$ is known for a very
regular subpulse drift in the so called B-mode. \citet{MTET13} found that the X-ray emission in the B-mode of PSR
B0943+10 is well described by thermal emission with blackbody temperature $T_s=3 \times 10^6$ K coming from a small hot
spot (with $R_{\text{BB}}$ of about 20 m, much smaller than the canonical polar cap radius $R_{\text{PC}}=140$ m) with
luminosity of $7 \times 10^{28}$ erg/s, in good agreement with predictions of PSG \citep{GMG03, GHMGZM08}. Indeed, for
$A_{\text{BB}}=1.4 \times 10^7$  cm$^2$ Eq. (1) gives the value of the surface magnetic field $B_s=1.7 \times 10^{14}$
G, much bigger than the dipolar surface component $B_d=4 \times 10^{12}$ G. The case of PSR B0943+10 is marked in Fig.
1 as the purple dot with co-ordinates $x=1.71 \times 10^{12}$ and $y=3.02$, which lies exactly at the theoretical line
of \citet{ML07}. Thus, the small size of the detected hot spot (the size of the actual polar cap) is naturally and
self-consistently explained by the magnetic flux conservation law expressed by Eq. (1). The ultra-strong surface
magnetic field required by the cohesive energy needed to form PSG in PSR B0943+10 can be indeed inferred from the
detected X-ray data. In this paper we argue that such strong magnetic field anomalies  can be naturally generated by
the Hall drift effect.

\section{The Electron-Positron Pair Creation}

The creation of a large enough number of electron-positron pairs is a basic prerequisite for models that claim to
explain the generation of the pulsar radio emission \cite[e.g.][]{RS75,AS79,GLM04}. Depending on the kind of radiation
that produces pairs with the bystander of the magnetic field, certain conditions have to be fulfilled. In the following
we describe in some detail processes of electron-positron pair creation and derive the requirements to be placed on
strength and curvature of the surface magnetic field $B_s$.

Basically, there are two kinds of high-energy $\gamma$ seed photons that can take part in the pair creation in the presence of
a strong magnetic field: the curvature radiation (CR) and the inverse Compton scattering (ICS) photons. Let us discuss the
CR photons first. There are two main parameters that affect the efficiency of CR pair creation process: the particle
(electron and/or positron) mean free path $l_p$ (before it emits a $\gamma$ photon) and the photon mean free path (before this
photon converts into a pair). The smaller these paths are the more efficient the pair creation process is.
According to Zhang et al. (1997)
\begin{equation}
l_p=l_{\text{CR}}\sim \frac{2\hbar c}{\gamma e^2} R_{\text{cur}} \,,
\end{equation}
where $R_{\text{cur}}$ is the curvature radius of magnetic field lines. Thus, while passing a distance $l$, the
particle can emit approximately $l/l_{\text{CR}}$ photons with energy $E_{\gamma}=h\nu_{\text{CR}} \approx 1.4 \times
10^{-17} \gamma^{3}/R_{\text{cur}}$ \citep[see][]{Jackson}. Part of the particle energy that is transferred into
radiation $\epsilon$ is given by

\begin{equation}
\epsilon =\frac{ E_{\gamma}}{\gamma m_e c^2}\frac{l}{l_{\text{CR}}}\approx 2 \times
10^{-13}\frac{\gamma^3}{R_{\text{cur}}^2}l\;. \label{epsilon}
\end{equation}

\noindent On the other hand, the photon with the energy $E_{\gamma} > 2mc^2$ propagating obliquely to magnetic field
lines can be absorbed by the field and as a result an electron-positron pair is created. Thus, the photon free path
$l_{\text{ph}}$ can be estimated as a mean distance which a photon should cover until it is absorbed by a magnetic
field. For strong magnetic fields \citep[$B\gtrsim 0.2 B_q$, see][for definition of $B_q$]{RS75} the photon mean free
path can be calculated as
\begin{equation}
l_{\text{ph}}\sim \frac{2mc^2}{E_{\gamma}} R_{\text{cur}}=1.2\times10^{11}\frac{R_{\text{cur}}^2}{\gamma^3}.
\label{lph}
\end{equation}

For typical pulsar magnetic fields the dipolar radius of curvature is of the order of $10^8$ cm. Thus, for Lorentz
factors $\gamma < 5\times10^6$ we get from Eqs. \ref{epsilon} and \ref{lph} that $\epsilon < 2.5\times 10^{-9} l$.
Therefore, even passing a distance of $l\sim R=10^6$ cm, where R is the NS radius, a particle will transfer much less
than $1 \%$ of its energy into curvature radiation. The photon free path $l_{\text{ph}}$ for the same parameters is
about $l_{\text{ph}}\sim 10^7$ cm, much too long to assure the effective pair creation inside the gap with height
smaller than 100 m.

To get a reasonable level of pair production, the radius of curvature has to be decreased by two or three orders of
magnitude, i.e. $R_{\text{cur}}\sim 10^{5 - 6}$cm, e.g. assuming $R_{\text{cur}}=5\times 10^5$cm, we find $\epsilon
\approx 10^{-4} l$ cm and $l_{\text{ph}}\sim2\times10^2$ (Eqs. \ref{epsilon} and \ref{lph}), i.e. the particle
transfers a substantial part of its energy while passing less than $100$ m (a gap height) and the emitted curvature
photons are converted into electron-positron pairs. Therefore, Eqs. \ref{epsilon} and \ref{lph} demonstrate that the
efficiency of the pair creation (i.e. $\epsilon$ and $l_{\text{ph}}$) is very sensitive to the curvature of magnetic
field lines and thus a small scale $B_s$ with a typical $R_{\text{cur}}\lesssim 10^6$cm is necessary for the efficient
pair creation, as already suggested by \citet{RS75}.

The resultant magnetic field at the polar cap is a vector sum of two components: star centered global dipole field and crust
anchored small scale field. In order to get the radius of curvature close to the surface sufficiently small ($<10^6$cm), the
small scale field has to be much stronger than the dipolar component close to the surface.

For the Inverse Compton Scattering (ICS) process calculation of the particle mean free path $l_{ICS}$ is not as simple
as for that of the curvature radiation. Although we can define $l_{ICS}$ in a same way as we have defined
$l_{\text{CR}}$, it is difficult to estimate a characteristic frequency of emitted photons. We have to take into
account photons of various frequencies with various incident angles.

One should expect two modes of ICS, resonant and thermal-peak. Under the gap conditions the cross-section of the
resonant ICS is much bigger then the cross-section of the thermal ICS. The resonant ICS happens if the photon frequency
in the particle rest frame equals to the electron cyclotron frequency, i.e. a magnetic field $B_s$ should be strong
enough to satisfy the resonant condition for X-ray photons. Detailed calculations demonstrate that again $B_s$ has to
be of the order of $10^{13} - 10^{14}$ G within a gap \cite[see][]{S13}.

\section{Processes that can Create Strong Small Scale Magnetic Field Structures
         at the Polar Cap}

As shown in the preceding sections, the 'magnetic spot' has to consist of strong and small scale, i.e. very curved,
components of the surface magnetic field. These structures have to be present for at least $\sim 10^6$yrs. These two
requirements have to be fulfilled in spite of the facts, that the ohmic lifetime of such small field structures is
significantly smaller than that of the global dipolar field which is maintained by currents circulating in the
highly conducting deep regions of the crust and/or in the perhaps superconducting core.

The lifetime of $B_s$ in the 'magnetic spot' is shortened due to its small scale and low electric conductivity.
The electric conductivity drops strongly towards the outermost shell of the crust because of the high density
gradient \cite[see Fig.1 in][]{PG07}.

Conventionally, the characteristic Ohmic decay time is given by

\begin{equation}
\tau_{\mathrm{Ohm}}=\frac{4\pi\sigma L^2}{c^2}
\label{tau-ohm-conv}
\end{equation}

\noindent where $\sigma$ denotes the scalar electric conductivity and $L$ is a typical scale length of the
field structure under  consideration. It is obvious, that an estimate of the magnetic field lifetime based
on Eq. (\ref{tau-ohm-conv}) is
not well suited to describe the evolution of  a such complex structure as present in the 'magnetic spot'.

More thorough analysis of poloidal field structure decay demands taking into account different field modes that form
the surface field configuration. If this is done by decomposition of that  structure into spherical harmonics, the
following decay times are found that are marked by the multipolarity $n$ \citep[see][]{GPZ00}:

\begin{equation}
\tau_{n}=\frac{4\sigma R^2}{\pi c^2 n^2}\;.
\label{tau-n}
\end{equation}

\noindent Here, $R$ denotes the radius of the sphere within which the field maintaining electric currents are
circulating and $\sigma$ is assumed to be homogenous throughout this sphere.

Obviously Eq. (\ref{tau-n}) is, given the constant conductivity assumption, still a rough approximation to the
situation in the 'magnetic spot' too. However, it takes the
complex and small scale structure of the field there better into account than Eq. (\ref{tau-ohm-conv}). Clearly, the
characteristic ohmic decay time $\tau_{\mathrm{0hm}}$ overestimates the duration of field decay even for a dipolar
field by about one order of magnitude. The typical scale length, i.e. the radius of dipolar field line curvature, for radio
pulsars is about $10^8$cm \citep{RS75}. As argued above, the typical scale length of $B_s$ has to be $\lesssim 10^6$cm.

The scale length of field structures expressed by spherical harmonics, that are determined by their multipolarity $n$,
is for small scales related to the radius of the sphere by $L=2\pi R/n$. For a neutron star radius $R \sim 10^6$cm and
a scale of $L\equiv R_{\text{cur}} \lesssim 10^6$cm, where $R_{\text{cur}}$ denotes the field line curvature radius of
$B_s$, returns $n \sim 10$, perhaps even significantly larger.

The electric conductivity in the layer below the surface of the 'magnetic spot' is much smaller than in somewhat deeper
crustal shells. This is caused both by the high temperature $T > T_s \gtrsim 10^6$K and by the low density $\rho
\lesssim 10^6$ g cm$^{-3}$. Extrapolating the results of \citet{C12} to that density range, the electric conductivity
will certainly not exceed $\sigma \sim 10^{22}$s$^{-1}$. Thus, for a field line curvature radius $R_{\text{cur}}
\lesssim 1$km corresponding to a multipolarity $n\approx 10$ this electric conductivity will cause according to Eq.
(\ref{tau-n}) an decay time of about $5000$ years. Hence, given the lifetime of radio pulsars, the small scale field
structure in the 'magnetic spot' has to be rebuilt continuously.

\subsection{Structures Created at a Neutron Star's Birth}

The short decay times for a poloidal field structure of $\lesssim 1$km scale length as estimated by use of Eq.
(\ref{tau-n}) rule out, that such magnetic spots, occasionally created during the birth of a neutron star by small
scale dynamo action, can survive over the radio pulsar lifetime. This is true the more, because during the early, say
$\sim 1000$yrs of a neutron star's life the crustal temperature is much higher which reduces the electric conductivity
additionally. Surface field anomalies created during the birth process of neutron stars can be responsible for the
radio pulsar activity of very young pulsars as e.g. the Crab-PSR. For pulsars older than $\sim 10^4$ years however,
these anomalies have to be re-created continuously.

\subsection{Polar Cap Currents}

A potential source of magnetic spots is the possible current through the polar cap. The Goldreich-Julian charge density
$n_{\mathrm{GJ}}\approx \Omega B_d/2\pi e c$. For a rotational period $P=2\pi/\Omega=1$ s and for $B_d=10^{13}$ G is
$n_{\mathrm{GJ}}\approx 7\times 10^{11}$ cm$^{-3}$.Thus, the maximum magnetospheric current density is
$j_{\mathrm{GJ}} e c \approx 10^{13}$ abamp cm$^{-2}$. Let's approximate Ampere's law by $B \approx (4\pi/c)j\times
L$. Then, for a scale length  $l\sim 10$ m the magnetic field strength  $\approx 4\times10^7$ G. Since the required
field strength at the magnetic spot should be $> 10^{13}$ G, even a much faster rotation and/or a larger scale of the
field would not be sufficient to generate by magnetospheric currents a field that can cause a significant curvature of
$B_d$ at the surface of the magnetic spot.

\subsection{Thermoelectric Magnetic Field Creation}

Somewhat less well understood is the role that the extreme heating of the small but heavily bombarded 'hot spot' for
the field evolution in the 'magnetic spot' can play. It is tiny in comparison to the whole surface and the heated layer
is very shallow. The surface temperature there exceeds certainly $10^6$ K while the large rest of the neutron star
surface has typically a temperature of $\sim 5\times 10^5$ K \citep[see][for the standard cooling history]{PGW06}.
However, due to the huge meridional temperature gradient a creation and perpetuation of small scale field structures is
conceivable. The battery term of the induction equation including thermoelectric effects, $\nabla Q\times\nabla T$,
where $Q$ is the predominantly radius dependent thermopower \citep[see for details ][]{BAH83, ULY86, GW91}, causes the
creation of a toroidal field structure. It surrounds the 'hot spot' as a torus, its inner radius corresponds roughly to
the scale of the meridional temperature gradient. First estimates show, that in this torus field strength of $\sim
10^{13}$ G can be created on timescales of $10^3$ years (Vigan\`{o} \& Pons, private communication 2013). However, this
field creation process and its further interaction with crustal field structures deserves much more detailed and
systematic studies.

\subsection{Hall Drift in the Crust}

The Hall drift in core and crust of neutron stars is subject of scientific debate since more then two decades. For the
very first time the term 'Hall drift' appeared in neutron star studies by \citet{BAH83}. They considered it as driving
force that can move a field around which has been created by a thermoelectric instability in the envelope. The first
who studied the effect of the Hall drift on the field decay in the crust was \citet{J88} . He argued that a non-zero
radial component of the Hall drift pushes magnetic flux, expelled from the superconducting core, below the neutron-drip
density where it will be rapidly dissipated by Ohmic diffusion.

Later \citet{HUY90}, \citet{YS91}, and \citet{US91} studied the influence of the Hall drift on to the Ohmic decay in
the core of neutron stars. For this purpose these authors calculated the Hall components of the conductivity tensor in
relaxation time approximation. These components have the same form both in the core and in the crust, the difference is
made by the different relaxation times for the diverse  collision processes.

Tensorial transport coefficients and the related Hall induction equation (see Eq. \ref{Hallind}) can only be used to
describe correctly the field evolution in the crystallized crust. There, the ions are fixed to the lattice sites and
the electrons are the only moving charged particles - a description in terms of electron-magnetohydrodynamics is the
correct one.

The more general case, when both electrons and ions (protons) can move with respect to each other and to the neutral
background of neutrons has been studied in detail by \citet{GR92}. In this pioneering paper they applied an (though
incomplete) analogy of the Hall induction equation with the vorticity equation of an incompressible liquid to discuss
the possibility, that the Hall drift causes the magnetic field to evolve in the crust through a turbulent cascade. It
should create out of large scale field modes increasingly smaller ones until their scale length becomes so small, that
Ohmic decay dominates the further evolution.

The realization that the thermal and magnetic evolution of neutron stars is closely related to each other is about 20
years old \citep[see e.g.][]{UM92, UR93, UCS94}. For the first time the magneto-thermal creation of small scale
toroidal field structures close to the North pole out of a large scale one via the Hall drift in the superfluid core of
neutron stars has been shown by \citet{US95}.

\citet{VCO00} studied for the first time the Hall drift in a medium, where a density - and, hence, conductivity -
gradient is present. Under such conditions the evolution of the toroidal field component follows a Burger's like
equation which describes the creation of current sheets in low density regions, i.e. sites of very efficient field
dissipation. However, even if the initial field configuration gives rise to a drift towards layers of higher
conductivity, the small scale of the locally intense field causes a much faster than Ohmic dissipation of magnetic
energy.

The occurrence of a Hall instability, i.e. the non-cascadic but jump-like, across wide spectral distances, transfer of
magnetic energy out of large scale into small scale modes has been studied by \citet{RG02} and \citet{GR02}. For the
first time the creation of small scale 'magnetic spots' at the surface of neutron stars via the Hall instability has
been demonstrated considering realistic crustal density profiles and cooling histories  by \citet{GRG03} and
\citet{RKG04}, however still in plan-parallel geometry. The occurrence of this instability has been disputed by
\citet{HR02} and \citet{HR04}, who extended the analytical studies of \citet{VCO00} to numerical ones, including the
simultaneous presence of toroidal and poloidal crustal field components. \citet{CAZ04} published a detailed and
comprehensive study of the crustal Hall drift. They found that a purely poloidal field is not affected by that drift if
its maintaining toroidal current consists of rigidly rotating electrons only; they confirmed the existence of the Hall
instability and discussed whether Hall waves can strain the crust beyond its yield point. In the same year \citet{J04b}
\citep[see also][]{J04a} explored the consequences of the existence of an amorphous phase (in contrast to an almost
homogeneous body-centered cubic crystal lattice) in deep crustal layers for the creation of small scale field
structures by the Hall drift. \citet{RBPAL07} studied  both analytically and numerically the evolution of axisymmetric
crustal field configurations. They showed that an exactly toroidal field is unstable to poloidal perturbations. At the
expense of the much stronger toroidal field, the Hall instability leads to the creation of strong and small scale
poloidal field structures close to the North pole \citep[see Fig. 2 in ][]{RBPAL07}. Note, that this resembles much to
what we assume for the 'initial' field configuration and to what is required to switch on a radio pulsar. A detailed
study of the crustal Hall drift for various 'initial' axial symmetric field configurations, realistic micro-physics,
and taking into account the cooling of the crust, has been presented by \citet{PG07}.

All numerical model calculations \citep{M94,NK94,SU97})  have been performed up to this time \cite[see also][]{KK12})
by using either a fully spectral code or a hybrid one, spectral in the angles and finite in the radial coordinate. The
obtained results demonstrate the non-linear coupling between different field modes, the exchange of magnetic energy
between them, and the acceleration of Ohmic diffusion. The latter occurs because the Hall drift, itself an energy
conserving process, creates tendencially an increasing number of field modes with small curvatures that are subject to
enhanced decay. These spectral methods, however, have very serious limitations on their applicability. Large gradients
of density and temperature, as well as in the electric conductivity and magnetization parameter cause problems with the
spectral resolution, especially at sites where current sheets in the crust are created.

To overcome this restrictions, in recent studies of the Hall drift another non-spectral but finite difference numerical
method was used. \citet{PG10} managed to show that the Hall instability indeed can appear in plan-parallel geometry. It
became clear that the field evolution during the Hall drift dominated phase proceeds not along a Hall cascade, but
rather through the rapid creation of very localized small scale structures in regions where the toroidal field changes
sign. In that way, possibly a Hall equilibrium will be established. For spherical and 3D symmetry however, such Hall
instabilities are expected to appear in a more complicated way because a larger variety of modes can become unstable
\citep[see][]{RBPAL07}.

A breakthrough in the treatment of Hall drift induced shock structures has been achieved by the development and
successful test of a new numerical code by the Alicante-group. A detailed and comprehensive study of this numerical
method is given by \citet{VPM12}. This code is not based on spectral methods but relies on upwind finite difference
techniques that are able to handle situations with very low, even vanishing, magnetic diffusivity and overcomes the
problems that are connected to the formation of sharp current sheets during the field evolution.

In the crystallized crust of neutron stars, where convective motions of the conductive material play no role, the
evolution of the magnetic field is governed by the Hall--induction equation

\begin{equation}
\frac{\partial\vec B}{\partial t}= - \frac{c^2}{4\pi}{\nabla \times}\left(\frac{1}{\sigma}
\left\{\nabla \times \vec{B} + \omega_B \tau [(\nabla \times \vec{B}) \times \vec b ]\right\} \right),
\label{Hallind}
\end{equation}

\noindent where $\vec b$ is the unit vector in the direction of the magnetic field $\vec b= \vec B/|\vec B|$, $\tau$ is
the relaxation time of the electrons and $\omega_B=eB/m_e c$ is the electron cyclotron frequency. Here, $\sigma$ is the
electrical conductivity parallel to the magnetic field (independent of the magnetic field strength).

The Hall drift term at the r.h.s. of Eq. (\ref{Hallind}) is a consequence of the tensorial properties of the
electric conductivity in the presence of a magnetic field. The tensor components of the electric conductivity are
derived in the relaxation time approximation \citep{YS91}.

If the {\it magnetization parameter} ($\omega_B\tau$) exceeds unity significantly, the Hall drift dominates the ohmic
diffusion and will result in a very different field evolution. A large magnetization parameter, typically $\approx
1000$ during some stages \citep[see Figs. 1 in ][and also Pons \& Geppert 2007]{GR02}, strongly suppresses the electric
conductivity perpendicular to the magnetic field and changes the character of Eq. (\ref{Hallind}). The magnetization
parameter

\begin{equation}
\omega_B\tau=\frac{e B(r,\theta,t)}{m_e^{\ast}(r) c}\, \tau[\rho(r),T(r,\theta,t),Q(r),A(r),Z(r)]\;
\label{omtau}
\end{equation}

\noindent depends on quantities that are in the present approximation functions of the radial coordinate only, as the
density $\rho$, the charge number $Z$, the mass number $A$, the impurity concentration $Q$, and the effective electron
mass $m_e^{\ast}$. Due to the certainly non-monopolar crustal magnetic field the heat flux through the crust will not
be spherically symmetric \citep{GKP04, PMP06}. Hence, the temperature $T$  depends on the radial and on the meridional
coordinate. Additionally, the magnetization parameter is linear in the magnetic field itself. This makes $\omega_B\tau$
to a spatially and temporally strongly varying quantity.

\subsection{Initial magnetic field configuration}

A basic feature of the Hall drift is the interaction between field modes that are different in their geometry -
toroidal or poloidal - and in their scale. The way how the Hall drift affects the field evolution depends strongly on
the initial field configuration. The meaning of 'initial' is not unambiguously. It is reasonable either to chose a
moment very close to the neutron star birth or the moment, when $\omega_B\tau$ exceeds unity. The importance of the
initial field geometry is demonstrated by the fact that, if there is initially only a purely toroidal field component
present, in axial symmetry never poloidal field modes will be created. In the opposite case, the sole existence of
poloidal modes in the initial field, the Hall drift will create out of them toroidal field components.

The precondition for the creation of the required 'magnetic spot' is the initial existence of a standard large scale,
i.e. dipolar poloidal field $B_d$, whose last open field lines determine the area of the canonical polar cap. For the
Hall drift to act efficiently, additionally a strong toroidal field $B_{\text{tor}}$ must be present too. The idea that
in the crust resides a very strong toroidal field is gaining more and more support \citep[see e.g. ][]{HG10, SL12,
VP12}. The open question is its structure. It has to fill almost the whole crust. Moreover, it has to reside in deeper
layers of the crust where the conductivity is high enough to guarantee a lifetime exceeding well $10^6$ years but not
too deep, where a hypothetical 'pasta'-phase with a huge impurity parameter can accelerate the toroidal field decay
\citep[see][]{J99, J01, PVR13}). That is to say that $B_{\text{tor}}$ has to be concentrated not only to a small
equatorial belt, as MHD equilibrium models suggest, but is strong as well in higher latitudes, closer to the poles. Our
preferred initial field configuration can have the form shown in the left panel of Fig.2. At the axial symmetric level
of field evolution modeling, initial toroidal field structures that are concentrated only in the equatorial plane do
not provide the required strong small scale field structures on acceptable time scales.

\begin{figure*}
    \centering
    \includegraphics[width=15.cm]{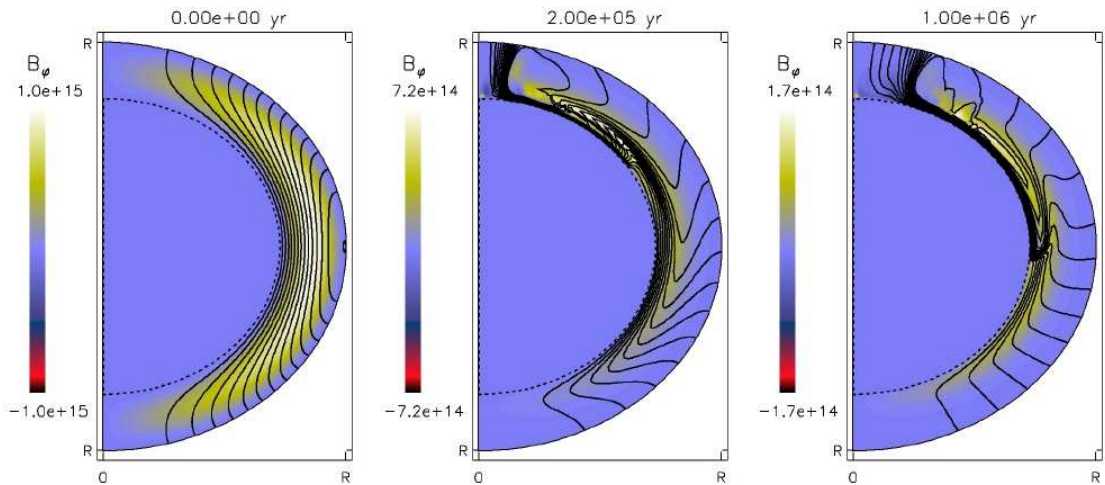}
    \caption{Structure of the crustal magnetic field at $t=0$ (left), after
        $2\times 10^5$ (middle) and after $10^6$ yrs (right). The poloidal
        field is shown by solid lines, the isolines of the toroidal field are
        color coded. The crustal region has been stretched by a factor of 4
        for visualization purposes. The complete movie showing the field
        dynamics is available at {\it http://personal.ua.es/en/daniele-vigano/hall-pulsar.html}.
                At the same address a movie cab be seen that shows the evolution of the external field lines;
                with and without the effect of the Hall drift too.}
\label{Bcrust_t}
\end{figure*}

It is still unclear whether an initial field configuration with an extended toroidal component actually exists in the
crusts of neutron stars which continue their life as radio pulsars. Studies of the MHD equilibrium in still liquid,
newborn neutron stars argue against it \citep{BS04,L10,GAL12,LJ12,FYE12}. An interesting attempt to conclude about the
internal field configuration by analyzing the magneto-elastic oscillation of neutron stars has been made recently by
\citet{GCFMS13}. In two very recent publications \citet{GC13a,GC13b} studied in axial symmetry the transition from MHD
equilibrium, established in barotropic matter, to a Hall equilibrium. Both equilibria are characterized by a toroidal
field that is confined to an equatorial belt, not extending to very high latitudes. There are, however, also
observational hints, that the crustal toroidal field has a structure similar to that assumed by us
\citep{RETIZSMTGGK10, SL12}). Small scale dynamos, as suggested by \citet{TD93}, could be a viable mechanism that
creates a toroidal field everywhere in the crust. Conceivable is also, that the submergence of a MHD equilibrium field
during the supernova fallback creates the required toroidal field \citep{GPZ99, H11,VP12,BPL13}. The question about the
initial field configuration will be certainly subject of the future scientific debate. However, the very existence of
radio pulsars should be taken into account then.

\section{Magnetic Field Evolution in the Crust}

As in the neutron star core, the magnetic field evolution in the crust cannot be considered detached from the thermal
history \citep[see e.g.][and references therein]{APM08a,PMG09,VRPPAM13}. Both quantities, temperature and magnetic
field, are intimately connected with each other there. The electric conductivity in the crystallized crust decreases
strongly with increasing temperature temperature \citep[see Fig. 1 in ][]{PG07}. Thus, the hotter the crust, the faster
the magnetic field decay. The magnetic field itself affects the cooling of the crust, either by channeling the heat
flow along the field lines and/or by being a Joule heat source. Since the Hall drift causes the creation of small scale
field structures, these are sites of enhanced Joule heating and sites of accelerated field decay too.

Physically, the Hall drift couples the field creating currents with the field itself in a non-linear manner. This
coupling creates in an energy conserving manner small scale field structures out of large scale ones. Without the Hall
drift, the initial magnetic field configuration would be stable, being subject only to Ohmic diffusion. Mathematically,
the Hall induction equation changes its character both spatially and temporally during the whole process of crustal
field evolution quite rapidly. In regions and in periods where Ohmic dissipation dominates, Eq. (\ref{Hallind}) is a
parabolic differential equation. However, where and when the magnetization parameter $\omega_B\tau \gg 1$, Eq.
(\ref{Hallind}) becomes a hyperbolic one. This behavior causes not only severe numerical problems. It makes also any
analytic approach to describe the total crustal field evolution e.g. in terms of ideal MHD questionable. A realistic
idea about the effects of the Hall drift on to the field evolution within the the crust, where steep density gradients
and a huge thermal inhomogeneity are present, can only be achieved by sophisticated numerical modeling.

The presently most advanced way to model the magneto-thermal evolution of the crust is to use the above mentioned code
of the Alicante-group \citep{VPM12}. A successful application of this code has been published recently by
\citet{VRPPAM13}. These authors were able to explain the observational diversity of isolated neutron stars in the
framework of the magneto-thermal evolution which is described just by that code.

In order to solve Eq. (\ref{Hallind}), the Alicante-group code has been applied. Since this code is especially able to
handle the specifics of the Hall drift it is well-suited to study the crustal field evolution during the first $10^6$
years of a neutron star's life. The 'initial' magnetic field structure consists of a poloidal dipolar field with a
polar surface strength of $10^{13}$ G. It has very little effect on to the development of 'magnetic spots' whether this
large scale poloidal field is anchored in the crust or is a core centered one. Important is the existence of a strong
toroidal field component that has to reside in deeper layers of the crust in order to survive a period of $\sim 10^6$
years almost undecayed. The 'initial' toroidal field shown in the left panel of Fig.2 has a maximum strength of
$10^{15}$ G.

In the middle and right panels of Fig.2 snapshots of the crustal field structure after $2\times 10^5$ and $10^6$ years
are shown. Obviously, strong and small scale surface field components are created close to the North pole and can be
maintained over $\sim 10^6$ years.  The axial symmetry is still conserved although there appears an extreme asymmetry
between the field structures near the North- and South-pole. This asymmetry is caused by the symmetry of the initial
magnetic field configuration adopted in our calculations (left panel). While the Hall drift velocity is proportional
to the field strength, the initial field symmetry determines a preferred Hall drift direction. It points towards one of
the poles depending on the relative sign of the toroidal field with respect to the poloidal one. In our case we have
chosen the negative signature that resulted in the drift towards the North pole. Of course, the opposite direction is
equally likely. However, we will argue later on, that the initial dipolar configuration is the only one that produces
the required magnetic spots on a proper timescale.

The role which the initial field configuration plays is best discussed by considering two qualitatively different
magnetic field configurations: dipolar and quadrupolar. As \citet{PG07} emphasized, there are two main effects that
determine the crustal field evolution, especially if there is initially a strong toroidal field component present. It
is globally displaced toward the inner crust following the negative conductivity gradient. Depending on the relative
sign with respect to the poloidal component it tends to move meridionally to one or the other magnetic pole. In case of
the initial toroidal field would have a geometry which is about quadrupolar, that motion would be North-South symmetric
and directed towards the equatorial plane.

For example, the evolution of a crustal field that consists of initially almost dipolar toroidal and poloidal
components is demonstrated in the second row of Fig. 6 in \citet{VRPPAM13}. The 'magnetic spot' has typically a
meridional extent of a few degrees which corresponds to a few hundred meters on the surface. A qualitatively different
situation is presented in Figs. 2 and 3 of \citet{PG07}, where the initial toroidal field adopted in calculations has a
quadrupolar structure but the poloidal component remains dipolar. In this case the Hall drift is equatorial symmetric
and directed from either pole towards the equatorial plane, creating there small scale structures. \footnote{One should
notice that the upper limit for the size of the 'magnetic spot' is the crust thickness which is about 1 km, consistent
with our numerical results mentioned just above. It is worth emphasizing here, that 'magnetic spot' and 'hot spot' are
different entities.}

Since the crustal conductivity profile is qualitatively similar for all isolated neutron stars, the location of the
'magnetic spot' depends crucially on the initial magnetic field geometry. Spots can be created at various latitudes,
provided the initial magnetic field has the corresponding structure. The term in the induction equation that describes
the meridional drift motion and the compression of field lines is of course the Hall drift term. The Ohmic diffusion
term describes the global trend of the toroidal field to migrate toward deeper crustal regions with higher electric
conductivity.

As it has been shown in Section 3, the quantities that are responsible for a sufficient creation of electron-positron
pairs are the surface field strength in the 'magnetic spot', $B_s$, and the radius of curvature of the field lines
there, $R_{\text{cur}}$, which are connected to the open dipolar lines along which the charges can be accelerated to
relativistic velocities within the polar gap. These quantities ($B_s$ and $R_{\text{cur}}=\left(-(\vec
b\cdot\nabla)\vec b\right)^{-1}$) are presented in Fig. 3. To show the striking effect of the Hall drift, the field
evolution as would appear without it (being subject only to Ohmic diffusion) is indicated by dashed lines. Without Hall
drift $B_s$ would decrease slowly from about $10^{13}$ G and the $R_{\text{cur}}$ would remain almost stable at a value
of a few hundred kilometers, a value typical for dipolar fields. The realistic situation is presented by solid lines.
For the  initial field strengths considered here (maximum toroidal field $10^{15}$ G, polar surface poloidal field
$10^{13}$ G) a 'magnetic spot' is created on a Hall time scale $\tau_{Hall} \sim  10^{3 - 4}$ years.

The Hall drift is a quite agile process that causes relatively rapid variations on the Hall time scale in that spot;
between a time step of $10000$ years the maximum of $B_s$ and the minimum of $R_{\text{cur}}$ varies by about $1^\circ$
within the spot. Since it is unimportant where within the 'magnetic spot' the strongest $B_s$ and the smallest
$R_{\text{cur}}$ are located at a given moment, both $B_s$ and the minimum of $R_{\text{cur}}$ are averaged over a
meridional range of about $5^\circ$, which is presented in Fig.3.

The results presented in Figs. 2 and 3 prove, that the for the functioning of radio pulsar emission necessary field
structures are created on a reasonable time scale: field strengths well exceeding $10^{13}$ G and curvature radii much
smaller than a few $100$ km. Moreover, an increase of the maximum strength of the initial toroidal field increases the
strength of the spot field too. Therefore, the resultant spot field can exceed  $10^{14}$ G for a longer period of
time. However, there is an upper limit for the crustal toroidal field strength. Indeed, an increase of the field is
associated with an increase of the Joule heating and when the strength reaches about $5\times 10^{15}$ G the Joule
heating leads to an enhanced Ohmic dissipation of the total crustal field.

The Hall drift tends always to create out of an initially large scale field a field of much finer structure. We have,
however, to emphasize, that the here presented initial crustal magnetic field geometry is the only one in the manifold
of examined ones that provides the required 'magnetic spot' within $\sim 10^4$ years surviving a typical pulsar
lifetime $\gtrsim 10^6$ years. It is likely that the number of favorable initial field structures increases when 3D
models can be calculated in the future.

\begin{figure}
    \centering
    \includegraphics[width=9cm]{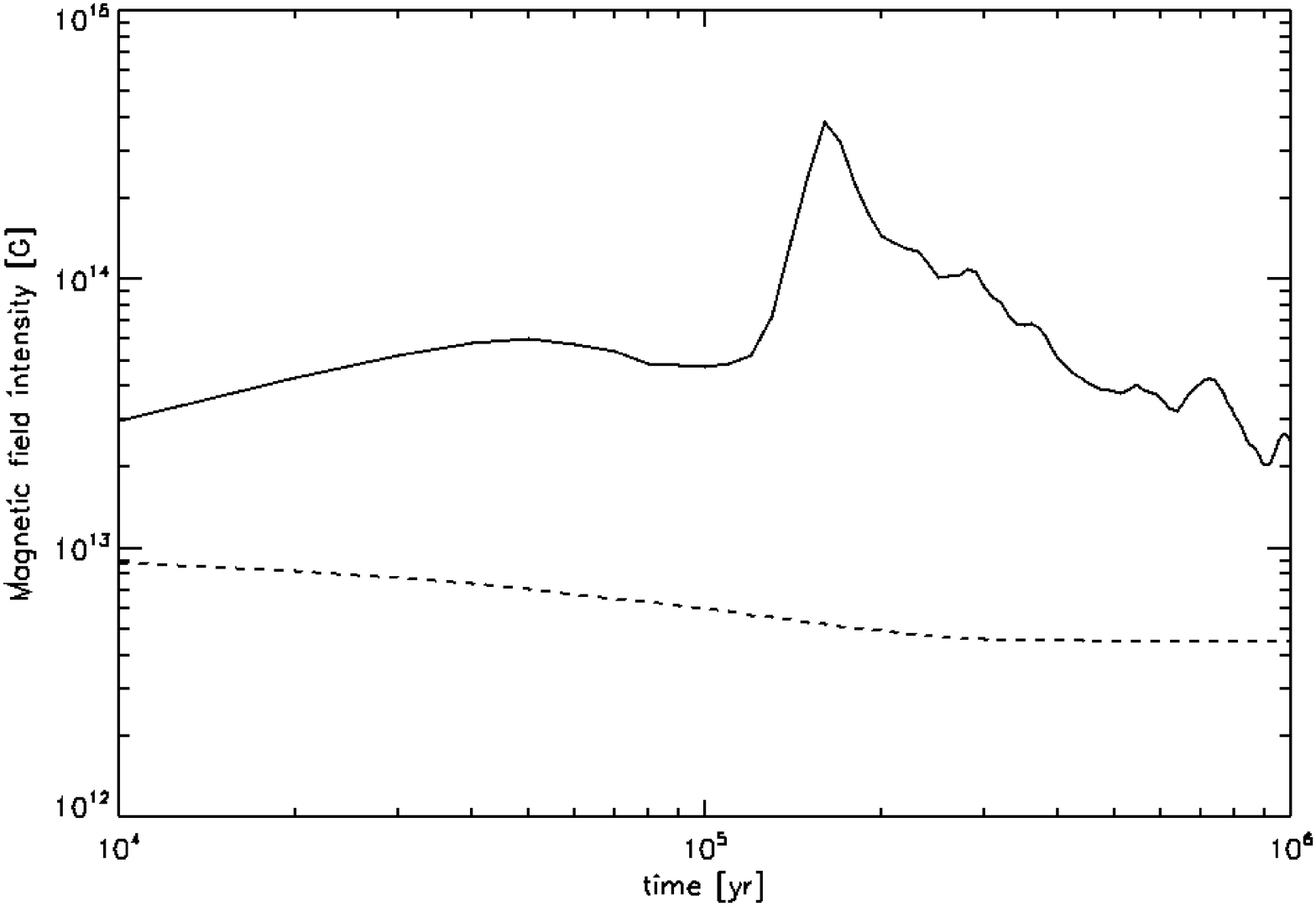}\\
    \includegraphics[width=9cm]{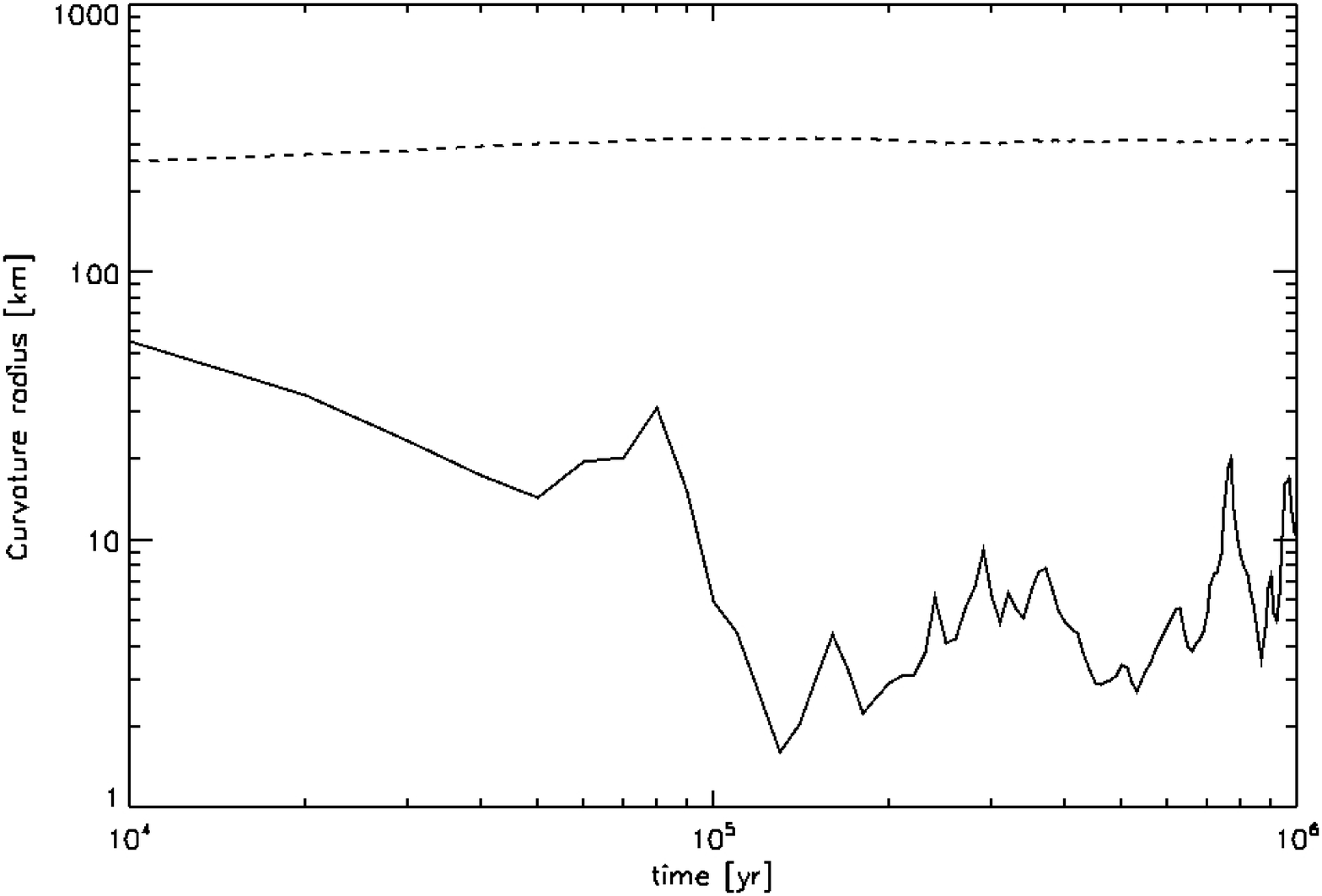}
    \caption{Temporal evolution of $B_s$ (upper panel) and $R_{\text{cur}}$ (lower panel) near the
    north pole, considering only Ohmic dissipation (dashed) or including the
    Hall term (solid). We show averages of the numerical values of $B_s$ and
    the minimum of $R_{\text{cur}}$ in the region $1^{\circ}-5^{\circ}$ from the
    north pole.}
    \label{B_RC_t}
\end{figure}

\section{Conclusions}

Our ideas and results can be comprised in the following chain of reasoning:\\
\begin{enumerate}
\item The brightness temperature of pulsar radio emission implies that a coherent radiation mechanism must be involved,
which requires creation of very dense, relativistic electron-positron plasma above the polar cap.
\item The inner pulsar acceleration region exists just above the polar cap, presumably in the form of PSG.
\item To create a large enough number of electron-positron pairs the curvature radius of surface magnetic field lines within PSG
has to be $\lesssim 10^6$ cm, i.e. the magnetic field structure has to be a 'magnetic spot' kind.
\item PSG can be created only if the cohesive energy of charges at the polar cap surface exceeds some critical value which is determined by the strength of the local magnetic field.
\item Combined radio and X-ray observations of several pulsars indicate that the actual polar cap, above which the acceleration arises, has a much smaller size than the canonical polar cap.
\item The surface temperature of the spot $T_s$ is of the order of $10^6$ K, which exceeds by far a typical surface temperature
corresponding to a cooling age of $\sim 10^6$ years. Thus, this temperature results from back flow bombardment of the
actual polar cap.
\item Both the small size of the actual polar cap (flux conservation arguments) and its high $T_s$ indicate
that the strength of the surface magnetic field $B_s$ is a of the order of $10^{14}$ G.
\item Such a magnetic field anomalies have to be maintained over a pulsar lifetime of $\sim 10^6 -  10^7$ years, at least in the polar cap region. Thus,
small scale field structures created at a neutron
star's birth are ruled out. The polar cap currents also can not be responsible for such a field.
\item To our knowledge, the only mechanism that can provide the strong, small  scale magnetic field structures is the Hall drift. It couples non-linearly
a standard large scale poloidal field ($10^{12} -  10^{13}$ G) and a
toroidal field that resides in deeper layers, occupying almost the
whole crust with a maximum strength of $\sim 10^{15}$ G.
\item Model calculations, although performed only in axisymmetric geometry, show, that the field structures necessary for the pulsar radio
emission can be created on the right time scale and with the required field strength via the Hall drift.\\
\end{enumerate}

We should now point out possible caveats of our idea. The first problem concerns location of the line that separates
"Gap" - "No Gap" regions in Fig. 1. This line describes dependence of the critical temperature \citep{ML07} on the
magnetic field strength.
Since \citet{ML07} calculated the binding energy for iron ions within an accuracy of a few tens of a per cent, this
curve might be slightly shifted towards the lower values for a given magnetic field strength. However, all calculations
are certainly correct in the sense of an order of magnitude estimates.

Also \citet{ZSP05}, \citet{KPG06} and \citet{PKWG09} pointed out, the black-body fits that return both $A_{\text{BB}}$
and $T_s$ are based on a small photon number statistics and have about the same quality as power law fits indicating a
magnetospheric origin of the radiation. Nevertheless, we believe that both the gap formation conditions and the
corresponding observational hints that these conditions are fulfilled are quite convincing to justify our assumptions.
On the other hand, one should realize that the pair creation process is also associated with the synchrotron
magnetospheric radiation, so the power law component must be present in the composite spectrum.

Another problem is the restriction of the model calculations to axial symmetry.  We expect that a future more realistic
3-dimensional treatment of the Hall effect in a neutron star's crust will provide an even better coincidence with
observations. Then, the 'magnetic spot' will occur as a real spot, restricted both in meridional and azimuthal
direction. This will make possible to describe the real shape and size of the actual polar gap.

\section*{Acknowledgments}
This paper is financed by the Grant DEC-2012/05/B/ST9/03924 of the Polish National Science Center. We acknowledge the
collaboration and many discussions with Jos\'{e} Pons and Daniele Vigan\`{o}. Their readiness to provide the numeric
results and  graphical representations made the completion of this manuscript feasible. We are also indebted to an
anonymous referee for exceptionally constructive criticism that helped us greatly to prepare the final version of the
paper.

 \label{lastpage}
\end{document}